\documentclass[12pt, times]{iopart}

\usepackage{iopams}  
\usepackage{graphicx}
\usepackage{xcolor}
\usepackage{wasysym}

\begin{document}

\title[Gaugain et al.: QSA error of electric field analysis for tACS]{Quasi-Static Approximation Error of Electric Field Analysis for Transcranial Current Stimulation}

\author{Gabriel Gaugain$^1$, Lorette Quéguiner$^1$, Marom Bikson$^2$, Ronan~Sauleau$^1$, Maxim~Zhadobov$^1$,  Julien Modolo$^3$, and~Denys~Nikolayev$^1$}

\address{$^1$ Univ Rennes, CNRS, IETR (Institut d'électronique et des technologies du numérique) – UMR 6164, 35000 Rennes, France.\\
        $^2$ Department of Biomedical Engineering, The City College of New York, CUNY, New York, United States of America\\
        $^3$ Univ Rennes, INSERM, LTSI (Laboratoire traitement du signal et de l'image) -- U1099, 35000 Rennes, France.}
         
\eads{ \mailto{gabriel.gaugain@univ-rennes1.fr} and \mailto{denys.nikolayev@deniq.com} }
\vspace{10pt}
\begin{indented}
\item[] December 2022
\end{indented}

\begin{abstract}

    \emph{Objective:} Numerical modeling of electric fields induced by transcranial alternating current stimulation (tACS) is currently a part of the standard procedure to predict and understand neural response.
    Quasi-static approximation for electric field calculations is generally applied to reduce the computational cost. 
    Here, we aimed to analyze and quantify the validity of the approximation over a broad frequency range. 
    \emph{Approach:}
    We performed electromagnetic modeling studies using an anatomical head models and considered approximations assuming either a purely ohmic medium (i.e., static formulation) or a lossy dielectric medium (quasi-static formulation). 
    The results were compared with the solution of Maxwell’s equations in the cases of harmonic and pulsed signals. Finally, we analyzed the effect of electrode positioning on these errors.
  \emph{ Main Results:} Our findings demonstrate that the quasi-static approximation is valid and produces a relative error below 1\% up to 1.43~MHz. 
  The largest error is introduced in the static case, where the error is over 1\% across the entire considered spectrum and as high as 20\% in the brain at 10~Hz. 
  We also highlight the special importance of considering the capacitive effect of tissues for pulsed waveforms, which prevents signal distortion induced by the purely ohmic approximation.
  At the neuron level, the results point a difference of sense electric field as high as 22\% at focusing point, impacting pyramidal cells firing times. 
  \emph{Significance:} Quasi-static approximation remains valid in the frequency range currently used for tACS. However, neglecting permittivity (static formulation) introduces significant error for both harmonic and non-harmonic signals. It points out that reliable low frequency dielectric data are needed for accurate tCS numerical modeling. 
  
\end{abstract}

\vspace{2pc}
\noindent{\it Keywords}: Electromagnetic dosimetry, finite element method (FEM), tissue dielectric properties, transcranial alternating current stimulation (tACS).

\submitto{\JNE}
%\maketitle

\section{Introduction}
\label{sec:introduction}
Transcranial current stimulation (tCS) is a non-invasive brain stimulation (NIBS) technique involving either direct (tDCS) or alternating currents (tACS), which are applied to the scalp with a fraction of the current reaching the cortex. 
 The interest about this technique is rapidly growing since tCS is a safe, cost-effective, and compact NIBS technology enabling home use with appropriate hardware \cite{bikson_safety_2016}. 
 Previous studies have suggested its potential to improve conditions related to several neurological disorders such as depression  \cite{bennabi_transcranial_2018}, stroke \cite{boggio_repeated_2007}, and Parkinson’s disease \cite{lee_does_2019}. 
The potential of tCS to enhance physiological cortical function has also been explored in healthy volunteers  \cite{nissim_effects_2019}.

The regain in popularity of tCS began in the 2000s with results showing that tCS increases cortical neurons excitability \cite{nitsche_excitability_2000}, which motivated the study of mechanisms involved at the cellular level. 
Pharmacological mechanisms have been studied, and significant changes induced by tDCS were demonstrated \cite{nitsche_catecholaminergic_2004, nitsche_pharmacological_2003, nitsche_gabaergic_2004}. 
Furthermore, electrophysiological studies have shown that the neuronal membrane depolarization induced by the exogenous electric field is proportional to the field magnitude  \cite{bikson_effects_2004}. This was supported by modeling studies with realistic cortical neurons \cite{rahman_cellular_2013}. 
The induced electric field magnitude in the brain is typically in the 0.1–1~V/m range for a standard protocol with a maximum intensity of 2 mA  corresponding on average to 0.12 mV per V/m of depolarization at the neuron level \cite{modolo_physiological_2018}. 
However, a membrane depolarization of the order of 20 mV is required to trigger an action potential, which is considerably higher as compared to the tCS-induced depolarization \cite{horvath_evidence_2015}. 
Some of the putative neuromodulation mechanisms include the modulation of the initiation timing of action potentials in the case of tDCS, and a facilitation of phase synchronization for tACS \cite{radman_spike_2007}. 
Initially, simple spherical head models have been used to provide a generalized view of tDCS mechanisms \cite{miranda_modeling_2006, datta_transcranial_2008} with a progressive shift towards more anatomically accurate shapes \cite{wagner_transcranial_2007}. 
Finally, various accurate MRI-based models of the head have been implemented \cite{datta_gyri-precise_2009, huang_new_2016}. 

Electric field distribution is generally computed numerically using, for instance, a finite element method (FEM) \cite{miranda_modeling_2006, datta_transcranial_2008, wagner_transcranial_2007, datta_gyri-precise_2009, opitz_how_2011}. 
The quasi-static approximation (QSA) -- assuming that the coupling between electric and magnetic fields is negligible -- is commonly used to model the induced electric fields of tCS \cite{ruffini_transcranial_2013}. 
In this approximation, there is no electromagnetic (EM) wave propagation. 
This is equivalent to the assumption that the wavelength is significantly larger as compared to the considered region size; therefore, the EM field phase variation is negligible across this region. 
This assumption is appropriate for tACS as it is mainly used at frequencies below 5 kHz \cite{chaieb_safety_2014} with free-space wavelengths in the order of 60 kilometers.
However, the guided wavelength inside a dielectric medium is inversely proportional to the square root of the relative permittivity, which can be as high as $10^6$ at this frequency for biological tissues \cite{gabriel_dielectric_1996, wagner_impact_2014}. 
This results in reduction of the wavelengths by a factor $10^3$ therefore affecting the range of validity of QSA. 
The second assumption is that electromagnetic induction can be neglected, which is valid since wave propagation effects can be ignored \cite{plonsey_considerations_1967}. 
The third commonly used assumption consists in neglecting the capacitive effect of tissues \cite{plonsey_considerations_1967}, i.e., considering biological tissues as purely ohmic \textcolor{black}{(i.e., neglecting the displacement electric field in Maxwell--Ampere’s equation). A forth hinted assumption, which is not often quoted, is to consider non-dispersive electrical properties (neither conductivity nor permittivity is time/frequency dependent).
These all four assumptions are those usually referred as quasi-static in the field of tCS and sometimes are suggested by referring to Laplace equation.}

\textcolor{black}{However, the last two assumptions are the most questionable ones~\cite{ruffini_transcranial_2013}}, since biological tissues \textcolor{black}{were shown} to have high relative permittivities – especially at low frequencies – and also strong dispersion \cite{foster_dielectric_1989}. 
\textcolor{black}{In theoretical and applied electromagnetics, the general QSA, also called electro-quasi-static, considers only the first two assumptions, which enforces to still solve the Laplace equation, but the three first assumptions together are equivalent to the static case (or quasi-stationary conduction) \cite{rapetti_quasi-static_2014,kruger_three_2019}. 
We will hereafter denote this case as "\textit{static}" approximation.}
In the case of the general QSA, the electrical properties of the dielectric medium act as a filter. The impedance becomes complex and therefore alters the shape of temporal waveforms \cite{wagner_impact_2014, bossetti_analysis_2008}. 

In the case of deep brain stimulation (DBS), this can affect the volume of tissue activated: an overestimation of about 18\% occurs considering only ohmic medium \cite{butson_tissue_2005}. 
The relative error of QSA in the electric potential analysis in the case of deep brain stimulation (DBS) is about 3\% to 16\% depending on the pulse duration \cite{bossetti_analysis_2008}. 
A point source in an infinite, homogeneous, and isotropic volume was used for the analysis in \cite{bossetti_analysis_2008}, and the general (full-wave) solution was compared to the static approximation.

Higher frequency spectra (and, therefore, shorter wavelengths) are being increasingly considered to improve the control of the fields induced in the head. Examples of such techniques include intersectional tDCS to reduce the heating of scalp tissues \cite{voroslakos_direct_2018} or temporal interference to target deeper brain regions using tACS \cite{grossman_noninvasive_2017, gaugain_embc2022, gaugain_bioem2022}. However, the approximation-induced computational errors are proportional to the operating frequency and can be significant \cite{gaugain_bioem2021}. To the best of our knowledge, no comprehensive error analysis has been performed for tCS in the case of heterogeneous realistic head models and realistic scenarios.

%Our previous work presented in conference introduced the present study \cite{} and a side study about the error introduced using purely conductive medium in the case of temporal interference was conducted \cite{, }. 

\textcolor{black}{In this study, we analyze and quantify the errors introduced by static and quasi-static approximations of tCS, as compared to the solution of Maxwell equations (full wave, denoted as FW hereafter). 
We first quantify the error induced by purely ohmic (i.e., static) and QSA approaches using 3D and 2D anatomical models of the human head for harmonic signals up to 100 MHz. The effect of uncertainty of low-frequency tissue properties on the computed error is considered next. In Section~\ref{sec_elec}, we assess the impact of electrodes placement. Time domain signals are considered in Section~\ref{sec_TD} for comparison with previous findings and for new stimulation protocols. Finally, we compute the impact of the error at the neuron level using biophysically realistic neuron.}

\section{Methods}

\subsection{Head model}
\bigskip
In order to evaluate the accuracy of the QSA, we first formulate a head model geometry and then numerically compute and compare the fields using both types of QSA and FW. 
The model geometry is based on the ICBM152 \cite{fonov_unbiased_2011} set of MRI segmented using the SimNIBS headreco routine \cite{thielscher_field_2015}. 
The resulting model consisted of five domains representing five main tissues commonly used to perform electric field modeling in a head: white matter (WM), grey matter (GM), cerebrospinal fluid, skull, and skin.
\textcolor{black}{All these domain are represented as a set of 3D surface meshes representing its outer boundary, forming the 3D geometrical model.}

\begin{figure*}[b!]
	\centering
		\includegraphics[width = \textwidth]{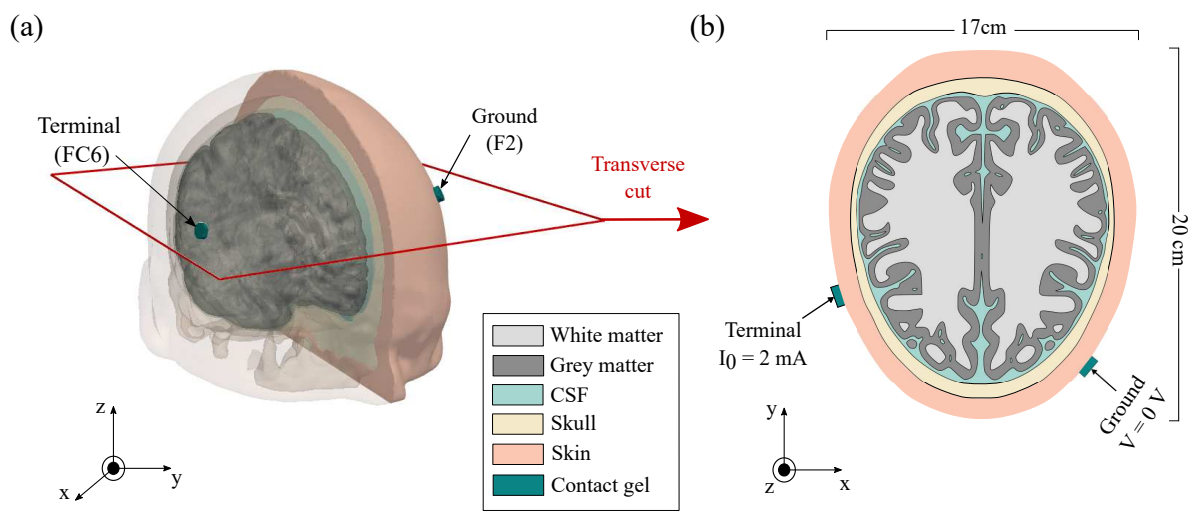}
	  \caption{Geometrical models of the head. (a): 3D model head model (with sagittal cut). The axial cut plane shown was used to build the 2D model.(b): 2D brain model including the segmented brain tissues. The tACS montage (position of electrodes and intensity applied at each electrode), dimension of the model and tissues modeled are illustrated.}\label{fig1} 
\end{figure*}

The 2D model was created using a slice from the segmented images was selected to represent the properly the geometrical complexity of the brain (gyri and sulci). 
Note that the final 2D model (\fref{fig1}b) should be seen as invariant by translation along the \textit{z}--axis. 
Clearly, it is a simplification of the human head that strongly varies along this dimension. 
However, this model has the advantage to enable the quantification of the relative error on the modeled electric field for different formulations, while also being computationally efficient. 
Since the QSA error is roughly a function of the ratio of the model dimensions \textit{a} to the wavelength $\lambda(\varepsilon)$ \cite{plonsey_considerations_1967}, the use of this simplified model for QSA error analysis is supported by the fact that the last dimension, along the \textit{z} axis, is theoretically infinite as aforementioned. 
\textcolor{black}{However the results has to be further validated with on the 3D model (see section 3.1.).}
 % Fig.~1 
%%%%%%%%%%%%%%%%%%%%%%%%%%%%%%%%%%%%%%%%%%%%%%%%%%%%%%%%%%%%%%

The two models were imported into COMSOL Multyphysics (COMSOL Inc., MA, USA), which was used for the field computation and error analysis.  
Two cylinders of 1-cm-diameters represented the contact gel for compact electrodes and were placed over the FC6 and F2 positions (\fref{fig1}a; placement according to the international EEG 10-20 system \cite{klem_ten-twenty_1999}). The electrodes were represented as semi-rectangular domains in the 2D model (\fref{fig1}b). 
Electrodes were modeled in terms of corresponding Dirichlet boundary conditions on exterior edge of the gel \cite{saturnino_importance_2015}.

\textcolor{black}{Finally, the surface mesh was generated using an L2 norm of error squared based mesh refinement in COMSOL Multiphysics, which lead to 125k triangular elements for the 2D model.
The 3D model was meshed with 5.31M tetrahedron elements. 
The smallest element edge was 0.5 mm and the largest one was 5 mm. 
This ensures convergence while having maximum element size much smaller than one tenth of the wavelength in each tissue for the highest considered frequency of 100~MHz.}

\subsection{Electric field modeling}
\bigskip

The electric field analysis requires a prior specification of the tissue dielectric properties [conductivity $\sigma$ (S/m) and relative permittivity $\varepsilon_{r}$]. 
\textcolor{black}{Since this study requires taking into account the dispersion due to the wide frequency range of interest}, we choose the established four-region Cole–Cole model with the coefficients tabulated by S. Gabriel and co-workers \cite{gabriel_dielectric_1996-1} since it i)~accounts for dispersive effects of tissues, ii)~allows to quantify the error introduced by neglecting the relative permittivity, iii)~satisfies the required Kramers--Kronig relationship \cite{bedard_generalized_2011}. 
The conductivity of the contact gel was set to 1.4~S/m \cite{datta_individualized_2011} and the relative permittivity to 80 as salt water. 
\textcolor{black}{Note that another important factor that might influence the estimated errors are the assumptions about the tissue electric properties. 
The low-frequency values (i.e., DC to tenth of kHz) found in the literature vary considerably \cite{mccann_variation_2019}, sometimes almost one order of magnitude, due to different conditions of the tissue and different ways of measuring. 
In particular, this set of conductivities is reported to deviate from literature in the low frequency range.
To analyze the effect of dielectric properties variation on the relative error, we performed the analysis at 10 Hz with some of the extreme cases reported in literature to indicate the range of electric field variation due to dielectric properties uncertainty.
}

The first formulation tested is the most used for tCS: the static formulation that neglects the propagating effect  \textcolor{black}{($\lambda \ll a$)} as well as the capacitive effect of tissues, i.e., the contribution of the relative permittivity ($\sigma \gg \omega \varepsilon_{r} $). 
The second is the quasi-static (QS) formulation, which only neglects the propagative effects, but not the permittivity contribution since the ratio between $\sigma$ and $\varepsilon_{r}$ (representing the dielectric relaxation time) is not negligible as compared to the typical variations of the electric field. 
This is also equivalent to considering a complex conductivity $\sigma_{c} = \sigma + j\omega \varepsilon_{r}$.  
The third and the most general formulation consists of solving the inhomogeneous  wave equation for the electric field, which is equivalent to solving the full set of Maxwell equations or full wave formulation (FW). 
 
For both static and QS formulations, the Laplace equation for the electric potential V [$\nabla \cdot (\sigma_c \nabla V) = 0$] was solved providing boundary conditions as follows: 

\begin{itemize}
    \item A Dirichlet boundary condition to model the ground (or cathode, $V=0$);
    \item A modified Dirichlet boundary condition (terminal boundary condition) on the anode, which imposes a constant current source ($\int \textbf{J} \cdot d\textbf{S} =0$) with a calculated fixed potential;
    \item An insulation boundary condition (Neumann) $\textbf{J} \cdot d\textbf{S} = 0$  on the remaining boundaries to model the skin–air interface. 
\end{itemize}

A stabilized formulation at low frequency (below 1 MHz) was used in FW computations, which is similar to the one described in \cite{zhao_new_2017, zhao_novel_2019} since common FW formulations are known to be unstable at low frequencies \cite{dyczij-edlinger_efficient_1999, jianfang_zhu_fast_2012}. 
The wave equation was decomposed into electric and magnetic vector potentials and solved on potentials rather than on the field directly. 
This formulation consists of solving Maxwell--Ampere's equation along with its divergence on electric and magnetic vector potentials, and appropriate boundary conditions as previously described supported with a Dirichlet boundary condition on the magnetic vector potential ($\textbf{A}\times \textbf{n} = 0$). 

The three formulations were solved on a mesh containing over 289k triangular elements for the 2D model and 5.31M of tetrahedrons elements for the 3D model. 
 MUMPS numerical solver was used to solve the linear system for the frequency range from 10 Hz to 100 MHz with 10 values per decade and with a relative tolerance of $10^{-6}$ for the 2D model. 
For the 3D model, appropriate iterative solvers formulation (Conjugate Gradient for static, BiCStab for QS and GMRES for FW) were used according to the formulation, with a relative tolerance of $10^{-6}$.  
\textcolor{black}{Finally, the relative error of the imposed approximation was computed using $\eta_{12} = \vert \vert \textbf{E}_1 - \textbf{E}_2 \vert \vert / \vert \vert\textbf{E}_1 \vert \vert  $  where 1 denotes the reference, being either FW when compared to the other formulations, or QS to compute the relative error between static and QS. }
The resulting error was computed over the whole numerical domain for each frequency, and the following metrics were computed: minimum, maximum, $2.5^{\mathrm{th}}$ quantile, $97.5^{\mathrm{th}}$ quantile, and mean. 

An additional study was performed to account for the electrode positioning. 
The skin contour curve, defined by the two coordinates (\textit{x}, \textit{y}), was interpolated according to the angle $\theta$ defined by the three following points: the fixed point in the frontal part of the head representing the cathode’s center, the center of the head and a third moving point on the contour. 
The latter represents the center of the anode which was moved to study the influence of the placement.
\textcolor{black}{See section 3.2 for the schematic of the setup.}

\subsection{Time domain waveform and harmonics}
\bigskip

Despite the typical use of sinusoidal signals in the case of tACS, temporal waveforms analysis might be useful for the elaboration of new \textcolor{black}{techniques} relying on waveform shaping to optimize the current delivery or even for shorter pulses used in intersectional tDCS (IS-tDCS) \cite{voroslakos_direct_2018}.
Once the electric field was computed for each formulation, the electric field values were exported from Lagrange’s points (vertices) of the mesh \cite{solin_higher-order_2003}. 
A post-processing routine was developed to convert these frequency-domain data into the time domain using Fourier series as:

\[ s(t) = \sum_n {c_n e^{2i\pi t f_{n}}  +c_{-n} e^{-2i\pi t f_{n} } }, \]
where $f_{n}$ is the frequency of the $n^{\mathrm{th}}$ harmonic and $c_{n}$ the associated Fourier’s \textcolor{black}{coefficient}. 
Fourier series were used to compute the electric field for typical time domain waveforms used for DBS, namely monophasic and biphasic pulses.
Pulse parameters were chosen in accordance with typical DBS waveform parameters: pulse duration was 90 µs, and the frequency was set to 130 Hz, which was comparable to the values used in \cite{bossetti_analysis_2008}\textcolor{black}{, with the same highest harmonic at 500 kHz and a sampling frequency of 1MHz}. 
Then, the relative error was computed in the time domain in the same way than in the frequency domain, for each time step between 0 to 400 µs.

\subsection{Impact on neuromodulation}
\bigskip

Electric field modeling during tACS is commonly \textcolor{black}{accompanied} by radial electric field calculation from the EF distribution \cite{rahman_cellular_2013}. 
This radial EF (EF component normal to the cortex surface) represents the EF along the pyramidal cells, which have a strongly preferential orientation normal to the cortex and are organized.
These cells showed the highest membrane polarisation due to the electric field with a direction parallel to their somato-dendritic orientation \cite{bikson_effects_2004}, which makes it a measure of tCS effect.
The radial electric field error was assessed similar to the previous relative error metric as  $\eta_{12} =  \vert \vert\textbf{E}_{r1}\vert - \vert \textbf{E}_{r2}\vert \vert  /  \vert\textbf{E}_{r1} \vert $.
The variation from the previous relative error formula was the difference of absolute values, i.e., the radial EF amplitude without taking into account the phase difference.
The impact of tACS was located at the cortex level where the field is the highest.
The 98$^{\mathrm{th}}$ EF quantile was computed over the cortical surface, and points with higher EF were selected to compute the radial relative error where accurate values are needed to predict the effect at the neuron level.
\textcolor{black}{We additionally assessed the same metric for the tangential component and the resulting angle between the electric field and the normal directions to the surface of the gray matter.
Finally, the phase difference between the different formulations was quantified since the phase term could have impact on phase activity \cite{johnson_dose-dependent_2020} and is supposed to not vary with location in the common static case.
}

\smallskip
To highlight the importance of these results, we performed neural modeling with a realistic neuronal model \cite{markram_reconstruction_2015} using the established NEURON software \cite{hines_neuron_1997}. 
Pyramidal cell from the 5$^{\mathrm{th}}$ cortical layer was used as it was demonstrated to be responsive to a 10 Hz tACS \cite{tran_effects_2022}.
The same mechanisms and setup were used as in \cite{tran_effects_2022}: a synaptic input was chosen to generate a 5Hz activity and the \emph{extracellular} mechanism was used to input the EF in the form of potential.
A 10 Hz tACS \textcolor{black}{was} used and values of EF were set using radial relative error results.
Three simulations were performed for three different EF amplitudes: 1.00 V/m as the reference, the additional average error on the radial field, and the maximum one.
Each simulation consisted of 140 seconds: 10 seconds of off stimulation and 2 minutes of on tACS and 10 seconds of off tACS.
Then, the phase-locking value (PLV) was computed to quantify the impact of tACS on neuron firing times, along with polar plots to quantify the timing influence of the stimulation.

%%%%%%%%%%%%%%%%%%%%%%%%%%%%%%%%%%%%%%%%%%%%%%%%%%%%%%%%%%%%%%%%%%%%%%%%%
\section{Results}   %%%%%%%%%%%%%%%%%%%%%%%%%%%%%%%%%%%%%%%%%%%%%%%%%%%%%
\subsection{Relative error spectrum}

\begin{figure*}[t!]
\centering
\includegraphics[width=\textwidth]{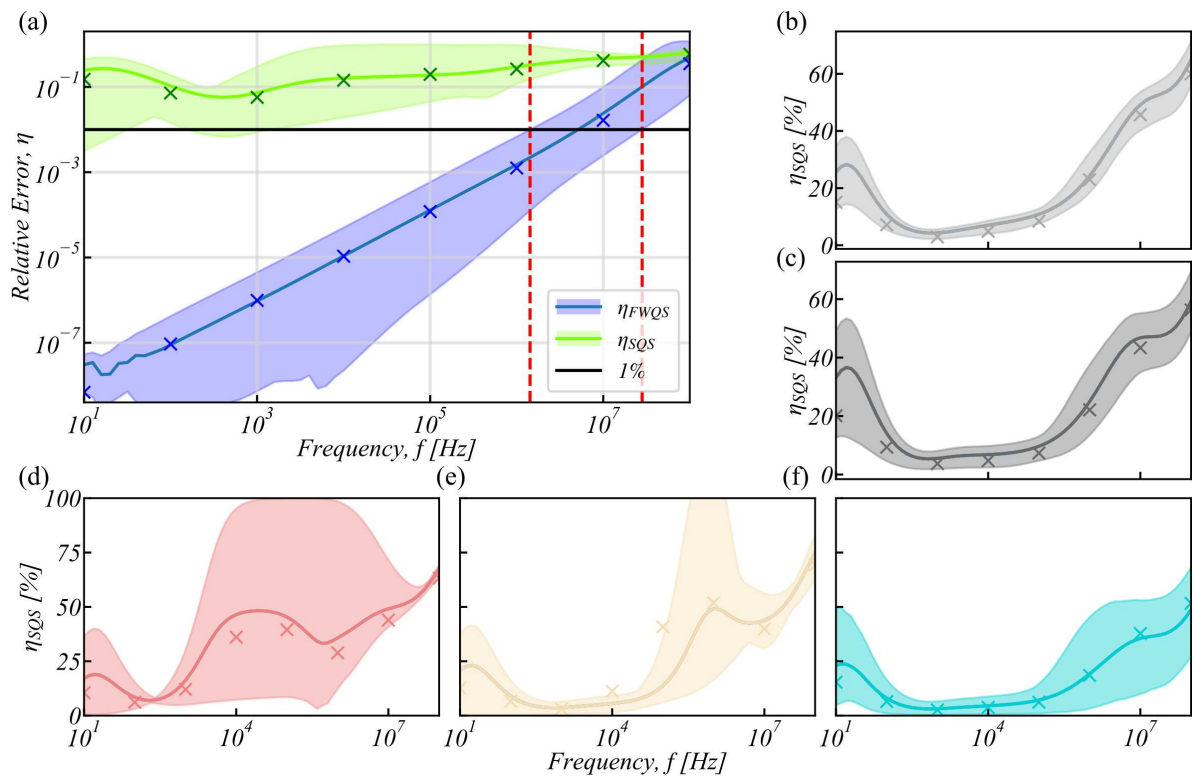}
\caption{ (a): Relative error spectrum between quasi-static and full wave approaches \textcolor{black}{($\eta_{\mathrm{FWQS}}$) and between static and quasi-static ($\eta_{\mathrm{SQS}}$), $f$ being the frequency}. The mean error\textcolor{black}{,} the $97.5^{\mathrm{th}}$ and $2.5^{\mathrm{th}}$ quantiles for both $\eta_{\mathrm{FWQS}}$ and $\eta_{\mathrm{SQS}}$ \textcolor{black}{with solid lines for the 2D model while the crosses represent the results for the 3D model}. The 1\% percent error line is shown and intersections between the $97.5^{\mathrm{th}}$ and $2.5^{\mathrm{th}}$ quantiles for $\eta_{\mathrm{FWQS}}$ is depicted as dotted red lines at 1.43 MHz and 28.16 MHz, respectively. 
(b) to (f): $\eta_{\mathrm{SQS}}$ in each tissue layer (in the order: WM, GM, skin, skull and CSF)  for the 2D model (solid lines and quantiles) and 3D model (crosses).}
\label{fig2}
\end{figure*}

Electric field maps were calculated over the considered frequency range, and the $97.5^{\mathrm{th}}$ and  $2.5^{\mathrm{th}}$ quantiles in addition to the mean relative error are illustrated in \fref{fig2}.
Both the relative error between FW and QS ($\eta_{\mathrm{FWQS}}$) and between static and QS ($\eta_{\mathrm{SQS}}$) are represented for the 2D and 3D models \textcolor{black}{for the FC6--F2 montage}. 
The relative error between FW and static, $\eta_{\mathrm{SFW}}$, is not shown because it is overlapping with $\eta_{\mathrm{SQS}}$ since  $\eta_{\mathrm{FWQS}} \ll \eta_{\mathrm{SQS}}$.
\textcolor{black}{The results for 2D and 3D models are in very good agreement, which validates the use of the simplified 2D model for the subsequent studies requiring prohibitive computations over the 3D mesh.}
The average of $\eta_{\mathrm{SQS}}$ was over 20\% in brain tissues within the frequency range of 10–40 Hz – a common range used for tACS since it corresponds with the frequencies of physiological brain oscillations (and so is $\eta_{\mathrm{SFW}}$).

In contrast, $\eta_{\mathrm{FWQS}}$ increases with frequency and crosses the 1\% error line in the MHz range. 
Table 1 summarizes 1\%, 5\% and 10\% limits for the multiple metrics described in the previous section. These metrics can be used to define the range of the QSA validity, depending on the error level that should not be exceeded.
%%%%%%%%%%%%%%%%%%%%%%%%%%%%%%%%%%%%%%%%%%%%%%%%%%%%%%%%%%%%%%%%%%%%%%%%%%%%%%%%%%%

%%%%% TABLE 1 !!!%
\begin{table}[h]
\caption{\label{tab1}Frequencies (MHz) at which the minimum, maximum, mean, 2.5\textsuperscript{th} and 97.5\textsuperscript{th} quantiles FW to QS relative error cross 1\%, 5\%, and 10\%, respectively } 
\begin{indented}
\item[]\begin{tabular}{c l l l l l }
\br
$\eta_{\mathrm{FWQS}}$ &  Min &  $q_{\mathrm{2.5}}$ & Avg & $q_{\mathrm{97.5}}$ & Max\\
 \br
 1\%  & $>100$ & 28.16 & 5.13  & 1.43  & 0.81\\
 5\%  & $>100$& 86.63 & 17.04& 5.92  & 3.74\\
 10\% & $>100$ & $>100$ & 27.97  & 10.04 & 6.71\\  
\br
\end{tabular}
\end{indented}
\end{table}
%%%%%%%%%%%%%%%%%%%%%%%%%%%%%%%%%%%%%%%%%%%%%%%%%%%%%%%%%%%%%%%%%%%%%%%%%%%%

\subsection{ \textcolor{black}{Effect of dielectric properties variability}}
\textcolor{black}{
The high relative error between static and quasi-static formulations is mainly due to the low values of electric conductivities of brain tissues in the used set of dielectric values.
To investigate the range of SQS relative error over the range of conductivities reported in the literature, we performed the simulation at 10 Hz with the extreme dielectric properties for GM, WM, skull, and skin. As the reported values of CSF do not vary significantly, we kept the conductivity and relative permittivity values at 1.654~[S/m] and 102, respectively, throughout this analysis).  
For GM, WM, skull, and skin, the range of conductivities was taken from \cite{mccann_variation_2019}.
On the other hand, the relative permittivities were set with reasonable extreme values due to the lack of litterature data on this parameter in the Hz range. 
All the considered values are presented in the table \ref{tab2}.
}

%%%%% TABLE 2 !!!%
\begin{table}[b!]
\caption{\label{tab2} \textcolor{black}{Extreme dielectric properties used to asses the range of SQS relative error at 10 Hz.}} 
\begin{indented}
\item[]\begin{tabular}{c c c c c c c c c}

\br
& $\sigma$ GM &  $\sigma$ WM  &  $\sigma$ Skull  & $\sigma$ Skin  & $\varepsilon_r$ GM  & $\varepsilon_r$ WM & $\varepsilon_r$ Skull & $\varepsilon_r$ Skin\\
\br
min & 0.06 & 0.0642 & 0.0182 & 0.137 & $10^5$ & $10^5$ & $10^3$ & $10^2$\\
max & 2.47 & 0.81 & 0.28 & 2.1 & $10^8$ & $10^8$ & $10^5$ & $10^4$ \\
\br
\end{tabular}
\end{indented}
\end{table}
%%%%%%%%%%%%%%%%%%%%%%%%%%%%%%%%%%%%%%%%%%%%%%%%%%%%%%%%%%%%%%%%%%%%%%%%%%%%

\begin{figure*}[h!]
\centering
\includegraphics[width=\textwidth]{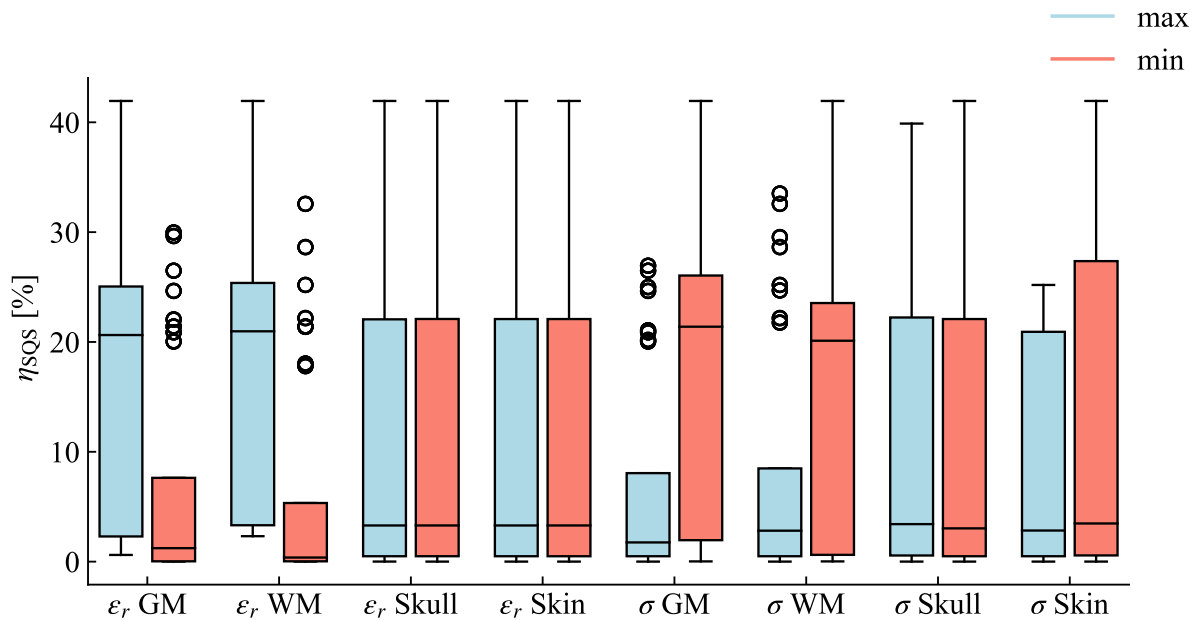}
\caption{ \textcolor{black}{Mean SQS Relative error at 10 Hz for all the extreme conductivity and relative permittivity values combination. The blue boxes represent the distribution for all the case where the maximum value of the property was used while the red correspond to the cases where the minimum value was used.} }
\label{fig3}
\end{figure*}

\textcolor{black}{
The distributions of the mean error, depicted \fref{fig3}, show higher errors for higher relative permittivity and lower conductivity, which is in agreement with the change in effective conductivity $\sigma_c = \sigma + j\omega\varepsilon$. 
It can also be observed that bimodal distributions occur when using minimum relative permittivities and maximum conductivities for brain tissues. 
This is mainly due to the fact that the ratio between $\omega\varepsilon_r$ and $\sigma$ is at maximum, which occurs when the conductivity and permittivity values are both at their minima or maxima. 
These results show a possible relative error at 10 Hz bounded between $2.10^{-3}$\% to 41.9\% depending on the considered properties and further confirms the need of reliable measurement of tissue dielectric properties in the sub-kHz frequency range. 
}

\subsection{Influence of electrode positioning and size}\label{sec_elec}

\begin{figure*}[!ht]
\centering
\includegraphics[width = \textwidth]{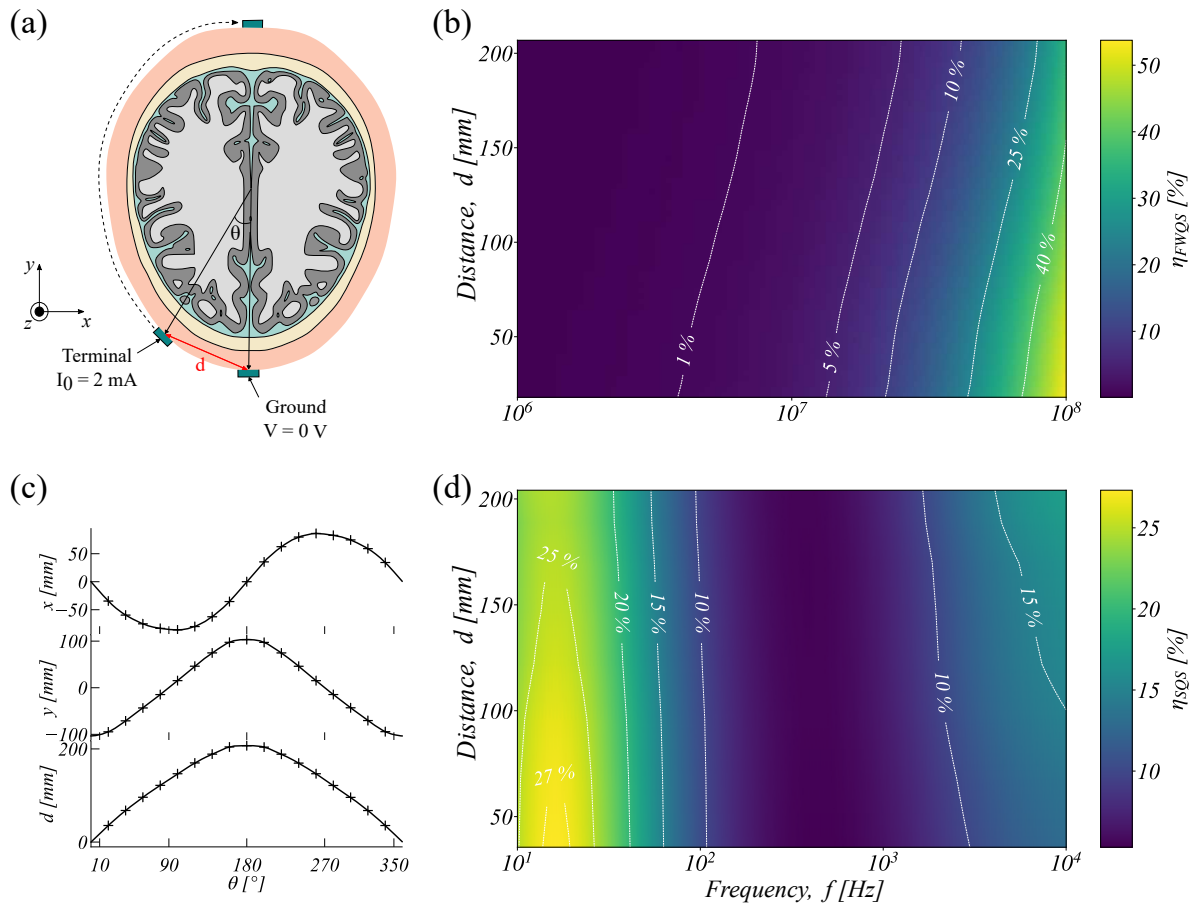}
\caption{ (a): 2D model with the angle defining the position of the anode by reference to the cathode position. 
(b): Relative error between FW and QS \textcolor{black}{($\eta_{\mathrm{FWQS}}$)} as a function of the distance \textcolor{black}{(d)} between anode and cathode in the 1--100 MHz range. The distance between the two electrodes is taken as the \textcolor{black}{vertical} axis, while the frequency \textcolor{black}{$f$} in log space in \textcolor{black}{horizontal} axis. 
(c):~\textit{x} and \textit{y} coordinates of the skin curve depending on theta and the associated euclidean distance. 
(d):~Relative error between static and QS \textcolor{black}{($\eta_{\mathrm{SQS}}$)} in the 10-Hz–10-kHz range, as illustrated with the previous plot.}
\label{fig4}
\end{figure*}

Next, we investigated the influence of the electrode montage on the approximation error. 
The relative error variation was quasi symmetrical with respect to the $\theta=180^{\circ}$ axis. 
This motivated to choose a parameter varying as symmetrically such as the euclidean distance between the spatial positions of the two scalp electrodes, denoted as \textit{d} (see \fref{fig4}a and 3c.  
Figure 3b depicts that the relative error between QS and FW decreases as the distance between the two electrodes increases. 

Conversely, $\eta_{\mathrm{SQS}}$ has non monotonic variations at low frequency (below 10 kHz). 
In the 10--100 Hz range, the error is higher with proximal electrodes but the effect is reversed in the kHz range as illustrated \fref{fig4}d.
The increase of $\eta_{\mathrm{SQS}}$ at the skin level in the kHz can explain this since the current is more distributed in skin when the electrode are more spaced.
Conversely, with proximal electrodes, the electric field is less distributed in skin and the error is more represented by the one in the GM at low frequency.

\textcolor{black}{This study considers $\diameter 1$~cm electrodes. In order to evaluate if the error is affected by the electrode size, we performed an additional analysis with $\diameter 2$ cm as it is another standard size for circular electrodes. No sensible variation in error was found, which indicates a negligible effect of the electrode size on QSA validity.
}

\subsection{Error for typical time-domain waveforms}\label{sec_TD}

% Fig.~5
\begin{figure*}[p]
\centering
\includegraphics[width=\textwidth]{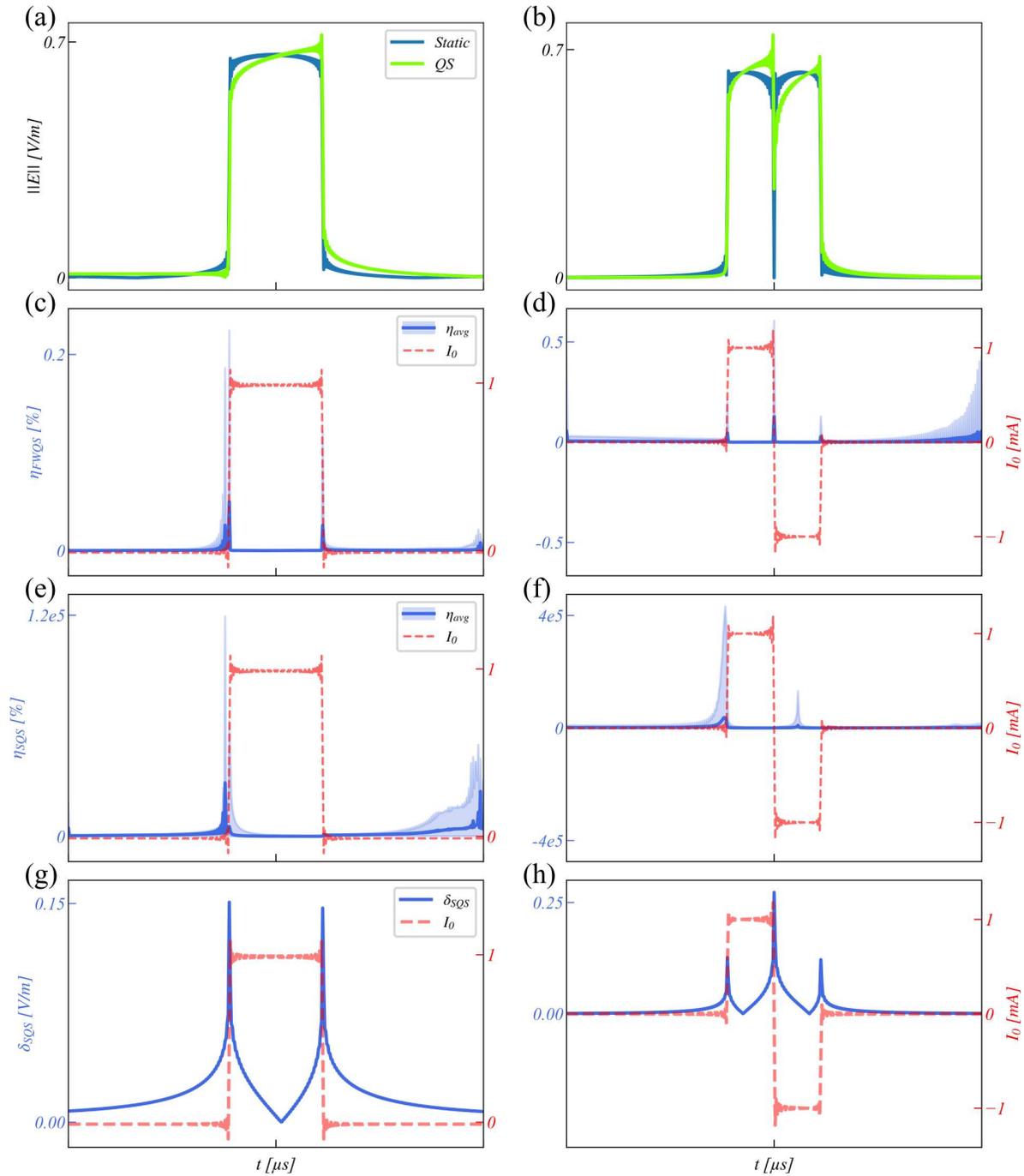}
\caption{(a)--(b): $97.5^{\mathrm{th}}$ highest electric field norm \textcolor{black}{($||E||$)} in grey matter for monophasic (a) and biphasic (b) pulses in Static and QS cases. 
(c)--(d): Average relative error between FW and QS \textcolor{black}{($\eta_{\mathrm{FWQS}}$)} in the time domain for (c) a monophasic pulse as a stimulus, and (d) a biphasic pulse, with the $2.5^{\mathrm{th}}$ to $97.5^{\mathrm{th}}$ quantile margin. The stimulus is represented in red with its corresponding second axis. 
(e)--(f): As (c)–(d) for the case of relative error between Static and QS \textcolor{black}{($\eta_{\mathrm{SQS}}$)}. 
(g)--(h): Norm of the difference of the $97.5^{\mathrm{th}}$ quantile electric fields \textcolor{black}{($\delta_{\mathrm{SQS}}$)} in the grey matter. }
\label{fig5}
\end{figure*}
%%%%%%%%%%%%%%%%%%%%%%%%%%%%%%%%%%%%%%%%%%%%%%%%%%%%%%%%%%%%%%

Using the Fourier’s series decomposition, the time domain relative error between QS and FW remained below 1\% for both square and biphasic pulses; \fref{fig5} demonstrates the general trends. 
The error was higher before and after the pulse with the highest values during the ascending and descending parts of the pulse, and smaller one during the positive phase of the pulse. 
This might originate from the difference in phase with the zero crossing of the finite harmonics signal. 
This is even more pronounced in the case of $\eta_{\mathrm{SQS}}$, which is tremendous due to zero crossings occurring at different times for Static and QS. 
Since it is mainly due to error in phase, it does not reflect properly the amplitude error, which is only represented during the positive phase of the pulse (and down state for the biphasic pulse), where the signal does not cross zero. 
This further justifies the choice made by Bossetti \textit{et al.} \cite{bossetti_analysis_2008} to represent the relative error only during positive state of the pulse. 
However, it does not highlight the error during at pulse termination which is substantial.  
% change here
Figure 5 shows that the results are in good agreement with \cite{bossetti_analysis_2008}, at least at the brain level where $\eta_{\mathrm{SQS}}$ \textcolor{black}{decrease from 14\%} during the first part of the pulse positive phase \textcolor{black}{while increasing during the second part and flare-up} at the pulse termination. 
This is mainly due to the zeros crossing of the pulse due to Gibb’s phenomenon. 
The norm of the difference between the compared electric field does not suffer from the aforementioned limitations and \textcolor{black}{quantify an error in electric field unit.} 
It is directly related to the amount of EF which is not present at the neuron level, and proportional to the membrane depolarisation. 
Figure.~4g and 4f illustrate this difference in electric field norm in the case of the $97.5^{\mathrm{th}}$ highest electric field, which is the zone where stimulation has the greatest impact. 
This difference is of the same order of magnitude than the electric field itself, which is significant and represents a difference of 22.7\% with the maximum value of the positive phase for a monophasic pulse (Static case) and 42.9\% for the biphasic pulse.

% Fig.~6
\begin{figure}[!ht]
\centering
\includegraphics[width=0.5\linewidth]{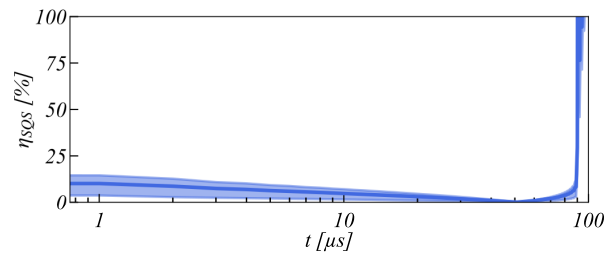}
\caption{SQS relative error \textcolor{black}{($\eta_{\mathrm{SQS}}$)} during the monophasic pulse up-state in the GM. \textcolor{black}{The horizontal axis is the time $t$ relative to the beginning of the pulse (0 being the start of the pulse)}. The mean value is plotted as a solid line and the margin represent the $97.5^{\mathrm{th}}$ and $2.5^{\mathrm{th}}$ quantiles.}
\label{fig6}
\end{figure}
%%%%%%%%%%%%%%%%%%%%%%%%%%%%%%%%%%%%%%%%%%%%%%%%%%%%%%%%%%%%%%

\subsection{Radial relative error}

\begin{figure*}[!hb]
	\centering
		\includegraphics[width = \textwidth]{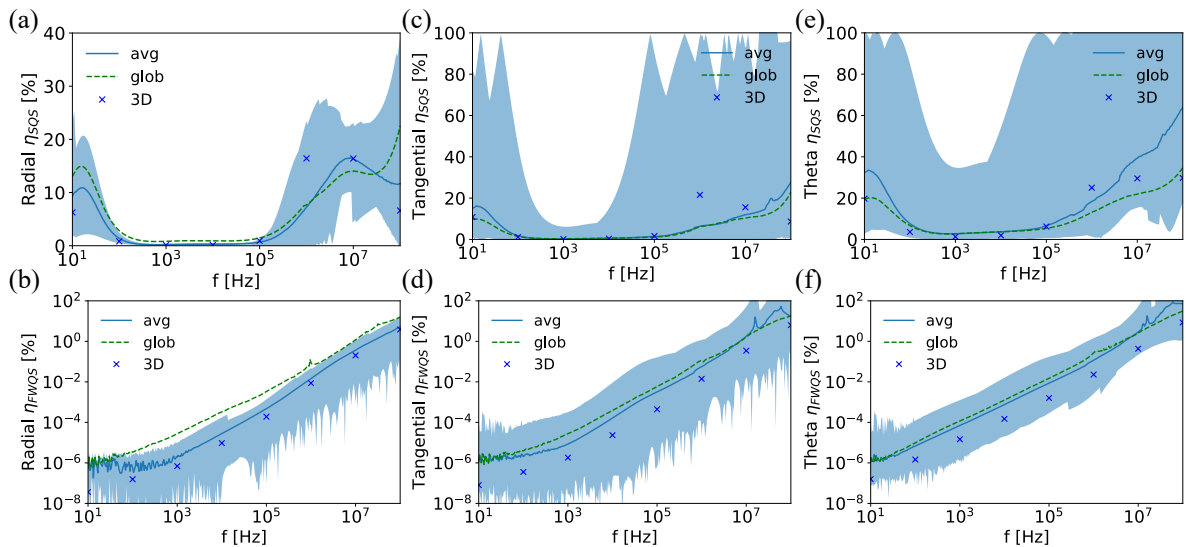}
	  \caption{Radial relative error between static and QS predictions (a) and between QS and FW (b). The continuous and dashed lines correspond to the 2D model results while the crosses represent the average radial relative error in the 3D model. The dashed green lines represent the average radial relative error over the full cortex. \textcolor{black}{ The same applies to tangential relative error for SQS (c) and FWQS (d) as well as for the angle SQS (e) and FWQS (f) relative errors}}\label{fig7} 
\end{figure*}

The radial\textcolor{black}{, tangential and angle} relative errors computed on the highest 2\% EF values over the cortical surface shows similar trends as over the full gray matter domain;
The results are presented in \fref{fig7} where the min--max margins over the 2D models and crosses for the 3D model are shown \textcolor{black}{for all the six resulting errors.} 
The curve for the average \textcolor{black}{relative errors} over the full cortex (all the EF values) \textcolor{black}{are} plotted as the green dashed line and is encompassed by the margins for $\eta_{\mathrm{SQS}}$ while it is slightly above the maximum in the case of \textcolor{black}{the FWQS radial relative error.}
At 10~Hz, which is a common frequency used for tACS \cite{tran_effects_2022}, the average radial relative error for the 2\% highest EF was about 6\% while the maximum reached 22\% in the case of SQS.
\textcolor{black}{The tangential relative error is higher with an even larger maximum in the full spectrum [\fref{fig7}(c) and (d)]. The average error (solid line) share same trend. Note that this higher maximum error can be due to the small absolute values of tangential field compared to the radial one, and relative error metrics are more sensitive to small field values. 
As a consequence, this impacts the error on the field orientation (angle between radial and tangential fields), which share the same trend.}
For the FW to QS, the average radial relative error remained below 1\% until 10 MHz, \textcolor{black}{while the tangential and angle relative errors cross the line at 7.24 and 5.01 MHz, respectively.}
% Fig.~7
%%%%%%%%%%%%%%%%%%%%%%%%%%%%%%%%%%%%%%%%%%%%%%%%%%%%%%%%%%%%%%

\textcolor{black}{
Finally, the phase error is depicted \fref{fig8} and shows the same trends as previous cases. 
To quantify the phase error, we use absolute absolute difference in radians between 1)~QS and static [SQS in \fref{fig8}(a)] and 2)~FW and QS [FWQS, \fref{fig8}(b)].
% Fig.~8
\begin{figure*}[!ht]
	\centering
		\includegraphics[width = \textwidth]{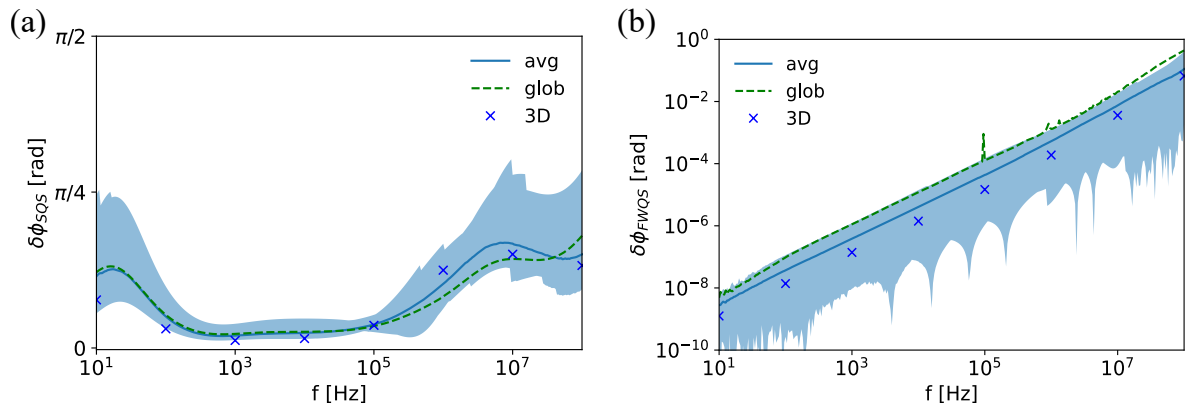}
	  \caption{\textcolor{black}{Phase difference (in radians) between static and QS, $\delta\phi_{\mathrm{SQS}}$ (a) and FW and QS, $\delta\phi_{\mathrm{FWQS}}$ (b). } }\label{fig8} 
\end{figure*}
%%%%%%%%%%%%%%%%%%%%%%%%%%%%%%%%%%%%%%%%%%%%%%%%%%%%%%%%%%%%%%
The average of SQS phase difference is up to $\pi/8$ in the 10--100~Hz frequency range. Its maximum is up to $\pi/4$, which represents a non-negligible phase difference from the neuromodulation point of view. Specifically, the neuron populations are stimulated with different phases depending on their location, which static approximation neglects.
However, in the FWQS case, the phase difference is quite negligible and increase log-linearly to cross a 1\% difference at 13.18 MHz.}

\subsection{Effect of tACS on single neuron activity}
\bigskip

\begin{figure*}[h!]
	\centering
		\includegraphics[width = \textwidth]{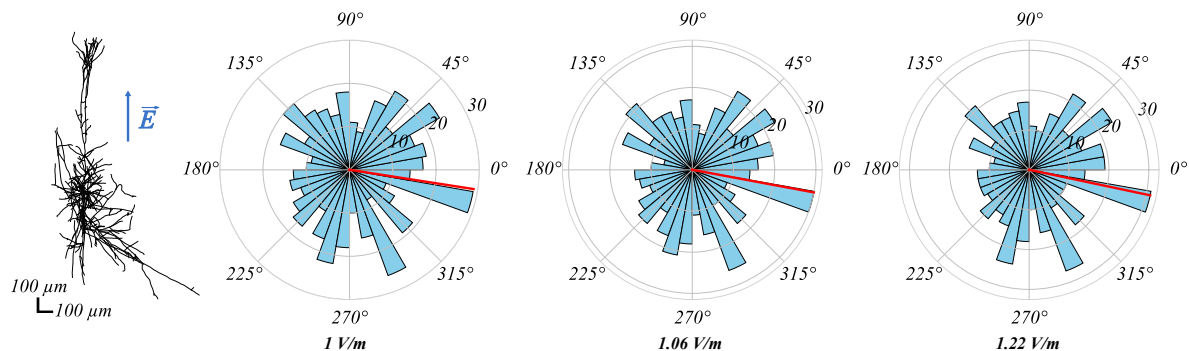}
	  \caption{Polar plots of the phase of spike occurrences distributions for the three different EF amplitude. The morphology is depicted in the left part with the direction of the input EF. Polar histograms correspond to event counts while the red line is the phase of the average vector. }\label{fig9} 
\end{figure*}

Using the previous results as an input for neural activity modeling of the selected pyramidal cell, the neural activity during tACS was computed with 1.00, 1.06 and 1.22~V/m.
All spike timing event were saved and then used to compute the distribution of spikes occurring in the same range of tACS waveform phase.
The corresponding polar plots are depicted in \fref{fig9} with the neuron morphology.
The distributions are close to each other since the sub-threshold input due to the extracellular field has little effect \cite{modolo_physiological_2018}.
However, the calculated PLV for each amplitudes are 0.0640, 0.0716, and 0.0798, respectively, which correspond to a 10.48\% increase in PLV for the average radial relative error and 19.66\% increase for the maximum one.
These results show the need of reliable EF predictions and the impact of taking into account the relative permittivity.

% Fig.~8
%%%%%%%%%%%%%%%%%%%%%%%%%%%%%%%%%%%%%%%%%%%%%%%%%%%%%%%%%%%%%%

\section{Discussion}

The major goal of this study was to assess the frequency-dependent accuracy of static and QS approximations commonly used in the tCS numerical analysis. 
We evaluated the tCS-induced electric fields in heterogeneous anatomical models for static, QS, and FW approximations. 
In terms of the error limits, the QSA 1\% error limit stands up to the MHz range exceeding 1\% at 5.16 MHz for the mean and at 1.43 MHz for the $97.5^{\mathrm{th}}$ quantile. 
This agrees well with the literature, where the limit at 1\% was identified using a plane wave illumination at 10 MHz \cite{park_calculation_2013}. 
In terms of the error between two possible QSA formulations – depending on whether one neglects the capacitive effect of tissues – we demonstrated, for the first time, that $\eta_{\mathrm{SQS}}$ is significant and even exceeds $\eta_{\mathrm{FWQS}}$ in the case of tCS. 
This is an important takeaway, since the inclusion of capacitive effects in the model does not significantly increase computational costs, especially as compared to a computationally expensive FW approach.

The FWQS relative error shows a linear-log increase over the frequency spectrum, as expected, since it is often quantified as being proportional to $\omega^2$ \cite{feynman_feynman_2011}, confirming the validity of QSA below the MHz range without neglecting capacitive effects. 
The interpretability of the SQS relative error is less straightforward, since it is mainly due to the change in the current distribution that is affected by the intrinsic impedance change. 
Note that in the low frequency range, in which tACS is currently performed (10–100~Hz), the SQS relative error is about 20\% for the 3D model, and it increases up to 50\% for the $97.5^{\mathrm{th}}$ quantile of the 2D model (figure.~2c) in the brain.
In high EF intensity areas, i.e. where brain is stimulated, this error can be as high as 22\% in the radial direction which therefore affects the firing times of pyramidal cells as demonstrated here. 
We hope that these results should encourage to consider the capacitive effect of tissues even at very low frequencies, since the relative permittivity is sufficiently high to induce significant errors in both amplitude and phase of induced electric field. 
Since EEG and tES are related by the reciprocity principle \cite{cancelli_simple_2016, dmochowski_optimal_2017}, EEG source localization methods could also be impacted by this error.
Currently, these methods are often formulated using purely ohmic tissues \cite{weinstein_lead-field_2000, michel_eeg_2019}. 
However, this frequency dependence would drastically increase the computational cost in this inverse problem.
It remains an open question how considering this frequency dependence of the permittivity would improve the performance of EEG source localization methods. 
The static approach might be still preferred for highly repetitive 3D modeling, such as the optimization of electrode placement \cite{saturnino_accessibility_2019}. 
In this case, an additional post-optimization QSA analysis might still be useful to provide more accurate values of electric field distribution.

The FWQS error was found to be a function of the distance between two electrodes, however limits remained within the same range (1–10 MHz for 1\% error, for example). 
The distance-error dependence also affected $\eta_{\mathrm{SQS}}$ at low frequencies. 
In the EEG spectrum domain (10--50Hz), the error decreased with distance, which can be explained by the higher error in the brain being more represented in the average one.
This even increased the error in the case of high definition tCS, where one electrode is closely surrounded by four others to increase focality of conventional tCS \cite{datta_gyri-precise_2009, edwards_physiological_2013}. 
This is a technique that is mainly used at low frequency (within the EEG frequency range: typically from DC to 100 Hz). Conversely, $\eta_{\mathrm{SQS}}$ increased as the electrodes were moved away in the frequency range used for the temporal interference technique (1--10 kHz).
This is mainly due to the increase of $\eta_{\mathrm{SQS}}$ in this frequency range in skin where the electric field is more distributed due to the electrode spacing.

Finally, the computed electric field in the Fourier space (frequency domain) can be transformed into the time domain and used to compute the corresponding relative errors.
Here, we presented two examples with 1) the monophasic pulse studied in \cite{bossetti_analysis_2008} for comparison and 2) the biphasic pulse that is a typical waveform used in brain stimulation and, in particular, for DBS \cite{de_jesus_square_2019}.
The results in the time domain suggest that the resulting error from using QS over FW was less than 1\%, validating the use of the QSA for this purpose. 
This level of numerical error is lower than 13\% reported by \cite{bossetti_analysis_2008} during the positive phase of the pulse. However, this difference is due to the comparison between the Static and FW formulations.
In our case, the error quantification showed a comparable range of error in grey matter supporting the rationale to include capacitive effects when the relative permittivity at low frequencies is high.
This supports the previous statements that neglecting the capacitive effect of tissues can be considered as an unreasonable approximation for most cases \cite{ruffini_transcranial_2013, bossetti_analysis_2008}.

This study addressed the question of the approximation for tCS electric field modeling in the case of a realistic head model with the main five tissues used in the literature.
The use of the Cole–Cole model can be criticized, since deviations in conductivity have been identified at low frequencies ($< 1$ MHz) \cite{gabriel_dielectric_1996-1}, which could be attributed to electrode-electrolyte interface during measurements \cite{gabriel_dielectric_1996, zimmermann_ambiguity_2021}.
This issue was recently addressed by compensating this electrode–electrolyte interface impedance \cite{zimmermann_ambiguity_2021}, which opens the possibility to use corrected values. 
However, another study reported similar range of values for relative permittivity but higher conductivities than the initial measurements, in mice tissues \cite{wagner_impact_2014} and is physically plausible.
Purely ohmic tissue models are plausible but singular due to Kramers–Kronig relations \cite{bedard_generalized_2011}. 
Still, in this model, skin has a conductivity of the order of $10^{-4}$ S/m, whereas it is commonly set in the 0.2–0.5 S/m range \cite{datta_gyri-precise_2009, wagner_three-dimensional_2004, geddes_specific_1967}. 
This could be explained by the fact that scalp tissues are multilayered, and composed of multiple tissues with their own properties, and that only surface skin was measured.
Yet, the conductivity used in Static and QS model are the same and we assess the QSA validity using a relative metric which is expected to be as high, even if more current is shunted through the scalp.
This illustrates further the need for reliable values of conductivity/permittivity at low frequency, where there is a large dispersion of values.
It is also worth to point out that most values were measured post-mortem, which can affect the results \cite{wagner_impact_2014}.
Another source of variability is inter-individual differences in brain morphology and conductivity \cite{huang_measurements_2017}, especially since such variability could be a larger source of error than these tackled approximations \cite{saturnino_principled_2019} and impact substantially the electric field distribution \cite{laakso_inter-subject_2015}. 
To overcome this limitations, we used a standardized (template) brain model, since the aim of this study was to show the intrinsic limitations of modeling practices, and the general tendencies of the error induced by the use of approximations, and not to extend exact values for every singular geometric model.
Finally, multiple electrodes stimulation montages could also be studied as an extension of the present study, since electrode positioning has been shown to have an important impact on the relative error distribution, especially comparing Static to QS.

\section{Conclusion}

\textcolor{black}{
This study provided an insight into modeling approximations commonly made in the research field of tCS. We demonstrated the validity of quasi-static approximation of Maxwell's equations until the MHz range if the relative permittivity is considered. However, the static approximation (i.e., purely resistive medium, no capacitive effects), introduces significant errors in tACS modeling in both the electric field amplitude and the phase. More importantly, static approximation assumes that the phase of induced electric field is the same across the brain. Our results demonstrate that  the phase can vary up to $\pi/4$ across the different regions of the brain, which is significant from the point of view of neuromodulation. Considering capacitive properties (i.e., relative permittivity of tissues, or, equivalently, the imaginary part of the conductivity) is especially important for pulsed signals that contain multiple frequency harmonics. 
Finally, precise knowledge of approximation-induced errors contributes to the better accuracy of computational modeling in tCS and therefore the analysis of associated effects at the cellular level.}

\section*{Acknowledgements}

This work has received a French government support granted to the CominLabs excellence laboratory and managed by the National Research Agency in the "Investing for the Future" program under reference ANR-10-LABX-07-01. 

\appendix
% Appendixes, if needed, appear before the acknowledgment.
\section*{Appendix: Maxwell's equations theory and approximations}
\setcounter{section}{1}

\textcolor{black}{
This appendix summarizes the equations for the electric field depending on the assumptions made in the article.
We start from the general set of Maxwell's equations to derive the ones used in approximations stated here as "static" and "quasi-static." The four Maxwell's equation can be written as:
}
\begin{eqnarray}
    \nabla \cdot \mathbf{D} = \rho  & \qquad ( \mathrm{Maxwell \! - \! Gauss} )  \label{eq1}\\
    \nabla \times \mathbf{E} = - \frac{\partial\mathbf{B}}{\partial t} & \qquad ( \mathrm{Maxwell \! - \! Faraday} ) \label{eq2}\\
    \nabla \cdot \mathbf{B} = 0 & \qquad ( \mathrm{Maxwell \! - \! Thomson} ) \label{eq3} \\
    \nabla \times \mathbf{H} = \mathbf{J} + \frac{\partial\mathbf{D}}{\partial t} & \qquad ( \mathrm{Maxwell \! - \! Ampere} ) \label{eq4}
\end{eqnarray}
\textcolor{black}{
where $\mathbf{D}$ is the electric displacement field,  $\mathbf{E}$ the associated electric field, $\mathbf{H} $ being the magnetic field, which is related to the magnetic flux density $\mathbf{B}$.
The conservation of the charge is obtained by taking the divergence of (\ref{eq4}):
}
\begin{eqnarray}
    \nabla \cdot \nabla\times\mathbf{H} = 0 = \nabla \cdot \mathbf{J} + \frac{\partial\rho}{\partial t}     \label{eq5}
\end{eqnarray}
\textcolor{black}{
Generally, the Maxwell's equations in time domain are computationally expensive to solve since the solution involves time convolution with electrical properties \cite{jackson_classical_1999}. 
In practice, the Fourier transform of the equations is computed since it involves a simple multiplications in the generalized Ohm's law $\mathbf{J} = \sigma\mathbf{E}$ and constitutive equations $\mathbf{D} = \varepsilon\mathbf{E}$ and $\mathbf{B}= \mu\mathbf{H}$ for linear media. 
This also simplifies the partial time derivatives, which are now simple multiplications with $j\omega$.   
In this study, we consider biological tissues that have a constant magnetic permeability $\mu = \mu_0$ but do not have constant relative permittivity. Considering that both $\sigma$ and $\varepsilon$ are functions of the frequency, the medium for which the equations are solved is \textit{dispersive}.
This is the most general case without any other assumption other than media linearity.
Therefore, considering (\ref{eq1}) and (\ref{eq4}), (\ref{eq5}) can be written as:
}
\begin{eqnarray}
    \nabla \times \mathbf{H} = \sigma\mathbf{E} + j\omega\varepsilon\mathbf{E} \label{eq6}\\
    \nabla \cdot \left[\left(\sigma + j\omega\varepsilon \right)\mathbf{E} \right]  = 0 \label{eq7}
\end{eqnarray}

\textcolor{black}{
The electro-quasistatic (EQS) approximation from the electromagnetics point of view consists in neglecting the effect of the induction on the electric field~\cite{rapetti_quasi-static_2014, kruger_three_2019}. This is reflected by the following change in (\ref{eq3}):}
\begin{eqnarray}
    \nabla \times \mathbf{E} = 0  \label{eq8}
\end{eqnarray}
\textcolor{black}{
It implies that \textbf{E} is a gradient of a scalar field, i.e., the usual relation to the scalar potential in quasi-static: $\mathbf{E} = -\nabla V$.
At this point, considering a dielectric medium, this results in the Laplace equation on the scalar potential by using (\ref{eq7}): }

\begin{eqnarray}
    \nabla \cdot [(\sigma + j\omega\varepsilon )\nabla V]  = 0 \label{eq9}
\end{eqnarray}
\textcolor{black}{
This simplification involves a spatial differential equation on a scalar quantity and is the equation solved for "quasi-static" case as referred to in the present work. 
In the neuromodulation research community, any form of Laplace equation is often cited as being the consequence of quasi-static assumptions. Additional assumptions are also considered to have an even simpler equation.
Indeed, often no dispersion is used (no frequency dependence of $\sigma$ and $\varepsilon$) together with neglecting the relative permittivity, i.e. assuming $ j\omega\varepsilon/\sigma \ll 1$.
This results in the Laplace equation on potential typically used in tACS numerical modeling:}

\begin{eqnarray}
    \nabla \cdot (\sigma \nabla V)  = 0 \label{eq10}
\end{eqnarray}
\textcolor{black}{
Eq.~(\ref{eq10}) formalizes what is commonly meant as "quasi-static assumption" in the neuromodulation community. Note that this contrasts with how EQS is defined in the electromagnetics/physics community, which often relates to this  equation as "static regime" or "quasi-stationary conduction"~\cite{rapetti_quasi-static_2014}. Since in this study we analyze the accuracy of both (\ref{eq9}) and (\ref{eq10}), we use the terms "static" and "quasi-static," respectively, in the text to distinguish these two approaches.}

\section*{References}
\bibliographystyle{unsrt}
\bibliography{mybib}

\end{document}